\documentclass[aps,pra,twocolumn,preprintnumbers,amsmath,amssymb,floatfix,superscriptaddress]{revtex4-1}

\usepackage{graphics}
\usepackage{graphicx}
\usepackage{amsmath}
\usepackage{setspace}
\usepackage{braket}
\usepackage{epstopdf}
\usepackage{comment}
\usepackage{hyperref}

\usepackage{bm}
\usepackage{xfrac}

\usepackage{xcolor}
\usepackage[normalem]{ulem}

\hypersetup{%
   pdfpagemode=None, 
   pdfstartpage=1,
   pdfmenubar=true,
   pdftoolbar=true,
   colorlinks = true,
   linkcolor=blue,
   citecolor=blue,
   urlcolor=blue,
   bookmarksopen=false
 }
\newcommand{\affA}{Van der Waals-Zeeman Institute, Institute of Physics, University of Amsterdam, 1098 XH
Amsterdam, The Netherlands}
\newcommand{\affB}{Faculty of Physics, University of Warsaw, Pasteura 5,
02-093 Warsaw, Poland}
\newcommand{\yb}[1]{$^{#1}\text{Yb}^+$}

\begin{document}

\title{Dynamics of a single ion spin impurity in a spin-polarized atomic bath}

\author{H.~F\"urst}\affiliation{\affA}
\author{T.~Feldker}\affiliation{\affA}
\author{N.~V.~Ewald}\affiliation{\affA}
\author{J.~Joger}\affiliation{\affA}
\author{M.~Tomza}\affiliation{\affB}
\author{R.~Gerritsma}\affiliation{\affA}

\date{\today}

\begin{abstract}
We report on observations of spin dynamics in single Yb$^+$ ions immersed in
  a cold cloud of spin-polarized $^6$Li atoms. This species combination has been
  proposed to be the most suitable system to reach the quantum regime in
  atom-ion experiments. For $^{174}$Yb$^+$, we find that the atomic bath
  polarizes the spin of the ion by 93(4)\,\%  after a few Langevin collisions,
  pointing to strong spin-exchange rates. For the hyperfine ground states of
  $^{171}$Yb$^+$, we also find strong rates towards spin polarization. However,
  relaxation towards the $F=0$ ground state occurs after 7.7(1.5) Langevin
  collisions. We investigate spin impurity atoms as possible source of apparent
  spin-relaxation leading us to interpret the observed spin-relaxation rates as
  an upper limit.  Using \textit{ab initio} electronic structure and quantum
  scattering calculations, we explain the observed rates and analyze their
  implications for the possible observation of Feshbach resonances between atoms
  and ions once the quantum regime is reached.
\end{abstract}
\maketitle
\section{Introduction}
\label{Sec_Intro}

In recent years, a novel field of physics and chemistry has developed in which
cold trapped ions and ultracold atomic gases are made to interact with each
other~\cite{Smith:2005, Grier:2009, Zipkes:2010, Schmid:2010, Zipkes:2010b,
Hall:2011, Hall:2012, Rellergert:2011, Sullivan:2012, Ratschbacher:2012,
Ravi:2012, Harter:2013, Hall:2013, Ratschbacher:2013, Haze:2015, Meir:2016,
Cote:2016, Tomza:2017cold, Sikorsky:2018, Haze:2018, Kleinbach:2018}.  These
efforts were motivated by the prospect of attaining ultra-cold
ions~\cite{Cot'e:2000a} by e.g.\ sympathetic cooling~\cite{Krych:2010,
Krych:2013, Meir:2016, Secker:2017} with atoms, probing atomic systems with
ions~\cite{Kollath:2007} and proposals to use the system for quantum
computation~\cite{Doerk:2010, Secker:2016} and quantum
simulation~\cite{Bissbort:2013}. However, these ideas require reaching
ultra-cold temperatures to enter the quantum regime which has proven very hard.
The main problem is posed by the time-dependent trapping field of the Paul trap
used to confine the ions. During a collision between an atom and an ion, energy
can be transferred from this field to the system, limiting attainable
temperatures~\cite{Zipkes:2010,Schmid:2010,Nguyen:2012,Cetina:2012,
Krych:2013,Chen:2014,Weckesser:2015,Meir:2016,Rouse:2017,Fuerst:2018}. In
reference~\cite{Cetina:2012}, it was calculated that the lowest temperatures may
be reached for the largest ion/atom mass ratios $m_\text{i}/m_\text{a}$. In this
work, we employ the ion/atom combination with the highest mass ratio of all
species that allow for straightforward laser cooling,
$m_\text{i}/m_\text{a}\approx 24\text{--}29$ given by Yb$^+$/Li. For this
combination, the quantum (or $s$-wave) regime is attained at a collision energy
of $k_{\rm B}\cdot8.6\,\mu$K, which should be in reach in state-of-the-art
setups~\cite{Fuerst:2018}.

For ion-atom mixtures to be used in quantum technology applications -  in which
quantum information will be stored in the internal states of the ions and atoms
- it is required that spin-changing collision rates are
small~\cite{Makarov:2003}. In a recent experiment Ratschbacher {\it et
al.}~\cite{Ratschbacher:2013} showed very fast spin dynamics in Yb$^+$
interacting with Rb atoms. Besides fast spin exchange - which conserves the
total spin of the collision partners - strong spin-nonconserving rates known as
spin-relaxation were observed. Very recently, spin-dynamics were also measured
in Sr$^+$/Rb~\cite{Sikorsky:2018}. Tscherbul {\it et al.}~\cite{Tscherbul:2016}
calculated that an exceptionally large  second-order spin-orbit coupling in
Yb$^+$/Rb provides a mechanism for the observed spin-relaxation rates. For
Yb$^+$/Li the second-order spin-orbit coupling is expected to be much
smaller~\cite{Tscherbul:2016}. A detailed knowledge about the spin-dependence in
cold atom-ion collisions gives insight into the possibilities of finding
magneto-molecular (Feshbach) resonances between the atoms and
ions~\cite{Julienne:2010,Idziaszek:2011,Tomza:2015,Gacesa:2017}. These play
a pivotal role in neutral atomic systems for tuning the atom-atom interactions
and find widespread application in studying atomic quantum many-body
systems~\cite{Julienne:2010,Bloch:2012}. In ion-atom mixtures, their existence
has been predicted~\cite{Idziaszek:2011,Tomza:2015}, but they have not been
observed so far since the required low temperatures have not been reached.
These considerations make an experimental study of the spin-dynamics in
Yb$^+$/Li of key interest.

 \begin{figure}[b]
  \begin{center}
    \includegraphics[width=0.9\columnwidth]{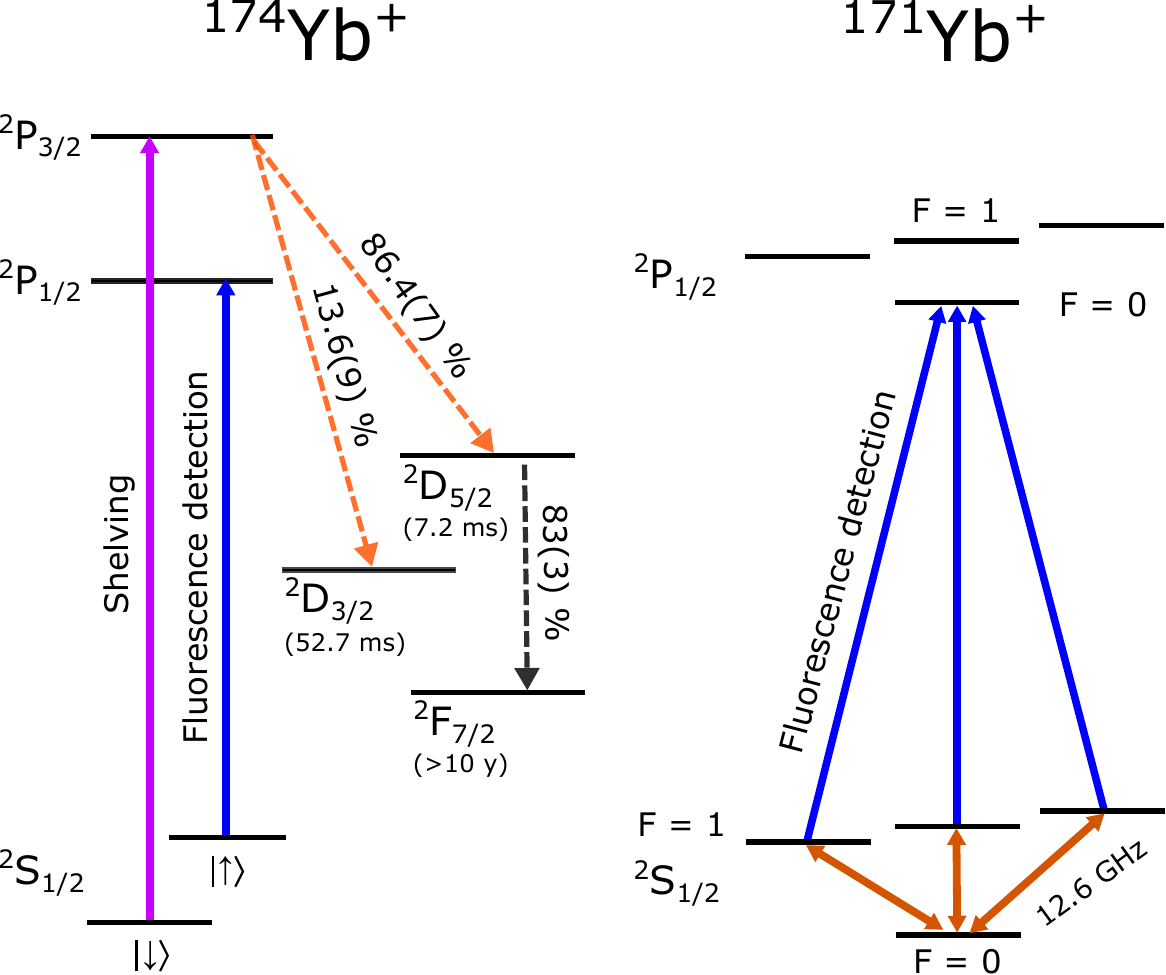}
  \end{center} \caption{Energy levels and relevant transitions in
   $^{174}$Yb$^+$ and $^{171}$Yb$^+$.}
  \label{fig:levels}
\end{figure}

In this work, we investigate the spin dynamics of single trapped Yb$^+$ ions in
a cold, spin-polarized bath of $^6$Li atoms. We prepare specific (pseudo-) spin
states in the ion by optical pumping and microwave pulses. Electron shelving and
fluo\-rescence detection allow us to determine the spin state after interacting
with the atomic cloud. For $^{174}$Yb$^+$ we find that the cloud of atoms
polarizes the spin of the ion by 93(4)\,\%.  Our results indicate a very large
spin-exchange rate of $1.03(12)\cdot\gamma_{\rm L}$, whereas spin-relaxation
rates are estimated to be $\leq 0.08(4)\cdot\gamma_{\rm L}$. Here, $\gamma_{\rm
L}=2\pi \rho_\text{Li}\sqrt{C_4/\mu}=22(7)$~s$^{-1}$ is the Langevin collision
rate, with $\rho_\text{Li}$ the density of Li atoms at the location of the ion,
$C_4$ is proportional to the polarizability of the atom and $\mu$ is the reduced
mass. For the $^{171}$Yb$^+$ isotope, we prepare all four hyperfine ground
states and measure all decay rates. As in $^{174}$Yb$^+$, we find strong rates
towards spin polarization. However, relaxation from the $m_F = 1$ state towards
the $F=0$ ground state occurs at a rate of $0.13(3)\cdot\gamma_{\rm L}$. All
relevant energy levels of both Yb$^+$ isotopes can be seen in
Fig.~\ref{fig:levels}. We combine {\it ab initio}\ quantum scattering
calculations with the measured spin dynamics. Interestingly, we
find in our calculations that even in the mK temperature regime the
spin-exchange rates still depend strongly on the difference between assumed
singlet ($a_\text{S}$) and triplet ($a_\text{T}$) scattering lengths. A similar
effect was observed in~\cite{Sikorsky:2018lock,Cote:2018}.
Our results indicate a large difference
between the singlet and triplet scattering lengths in Yb$^+$/$^6$Li, which will
be beneficial for the observation of Feshbach resonances. Our electronic
structure calculations also confirm that spin-nonconserving relaxation rates due
to second-order spin-orbit coupling should be smaller than for
Yb$^+$/Rb~\cite{Tscherbul:2016}.

\section{Experiment}

\subsection{Setup}
The experimental setup has been described in detail in Ref.~\cite{Joger:2017}.
In short, a cloud of magnetically trapped $^6$Li atoms in the $^2S_{1/2}$
$\ket{F=3/2,m_F=3/2}$ electronic ground state is prepared $2.1\,$cm below the
ion. The atoms are transported towards the ion by adiabatically changing the
magnetic field minimum position to a position $150\,\mu$m below the ion, where
the $\sim$~400~$\mu$m wide cloud interacts with the ion for a period of time
$t_\text{int}$. Afterwards, the atoms are transported back, released from the
trap and imaged on a CCD camera. Approximately 7$\cdot$10$^6$ atoms interact
with the trapped ion at a peak density of 49(15)$\cdot$10$^{14}\,$m$^{-3}$ and
a temperature of $T_{\rm a}=0.6(2)\,$mK.

The energy of the ion is composed of its micromotion in the Paul trap and its
secular energy. Before the experiment, we measured and compensated the ion's
excess micromotion in all three dimensions as described in
Ref.~\cite{Joger:2017}. We estimate a residual excess micromotion energy of
$\approx 2$\,mK per direction. We employ microwave sideband spectroscopy on
a single $^{171}$Yb$^+$ ion~\cite{Mintert:2001,Ospelkaus:2011} and infer an ion
temperature in the secular motion of $\approx$~4~mK after Doppler cooling and
a heating rate of less than 4\,mK/s. The combined energy of the ion is $E_{\rm
Yb}/k_{\rm B} \leq 20$\,mK during the experiments. Due to the large mass ratio
$m_\text{i}/m_\text{a}$, however, the collision energy $E_{\rm col}
= \frac{\mu}{m_{\rm Yb}} E_{\rm Yb} + \frac{\mu}{m_{\rm Li}} E_{\rm Li} \approx
k_\text{B}\cdot 1\,$mK is dominated by the energy of the atoms.

During the interaction, the ion experiences a magnetic field of 0.42\,mT caused
by the magnetic trap. The energy splitting of the ion's magnetic sublevels is
therefore kept small, allowing for spin-exchange.  Following each experimental
run, control measurements are performed to verify the conservation of the ionic
spin in the sequence when no atoms are loaded.

\subsection{Spin preparation and detection}

\begin{figure}[ht]
  \begin{center}
    \includegraphics[width=0.9\columnwidth]{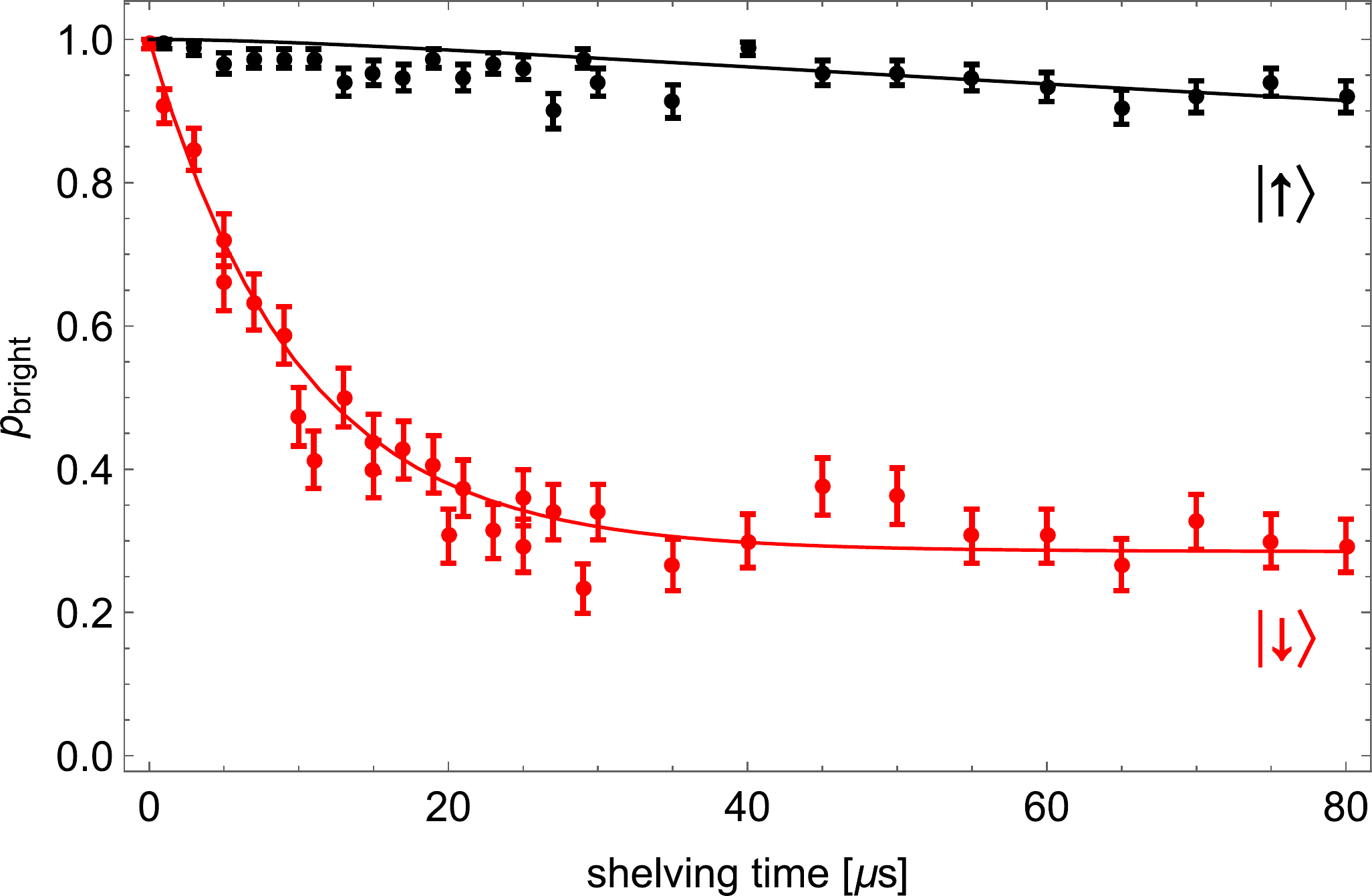}
  \end{center} \caption{Probability for finding an ion in the bright state
  versus duration of the 329\,nm shelving pulse. When initially prepared
  in the $\ket{\downarrow}$ state (red/gray) the ion remains unshelved with a
  probability of $28(3)\,\%$, whereas in the case of $\ket{\uparrow}$ (black) the
  ion is only shelved off-resonantly. The data is shown along with an analytic
  model that involves all relevant levels.}
  \label{fig:shelving}
\end{figure}

To initialize the $^{174}$Yb$^+$ ion in a Zeeman level of the $^2S_{1/2}$ ground
state, we apply a pulse of resonant circularly polarized light on the $369\,$nm
cooling transition along the trap axis. A small magnetic bias field pointing
either parallel or anti-parallel along the trap axis is used to prepare each of
the two Zeeman states.  We measure the optical pumping efficiency by comparing
the fluorescence during the optical pumping pulses for the correct
$\sigma$-polarization and the fluorescence for linear polarization with the
case where no ion is present.
From these measurements we obtain an optical pumping efficiency of $98.5(6)\,\%$
for the $\ket{^2S_{1/2},m_J=1/2}=\ket{\uparrow}$ state and $97.8(7)\,\%$ for the
$\ket{^2S_{1/2},m_J=-1/2}=\ket{\downarrow}$ state.

To detect the spin state after the interaction with the cloud of atoms, we
state-selectively shelve the ion into the long-lived $^2F_{7/2}$ state as
sketched in Fig.~\ref{fig:levels}~(left). We therefore apply a homogeneous
magnetic field of $72.5$\,mT to separate the ${^2S_{1/2}} \rightarrow
{^2P_{3/2}}$ transitions by 680\,MHz and irradiate
a shelving pulse resonant with the
$\ket{\downarrow}\rightarrow\ket{^2P_{3/2},m_J=-3/2}$ transition, allowing for
a decay channel via $^2D_{5/2}$ to the $^2F_{7/2}$ state with a probability of
$72(3)\,\%$~\cite{Taylor:1997,Biemont:1998,Feldker:2017}.  We measure the
probability for finding the ion being still in the $^2S_{1/2}$ ground state by
switching off the magnetic field after the shelving and subsequent detection of
the fluorescence during Doppler cooling.  The remaining population in the
$^2S_{1/2}$ state then contains both the unshelved and the imperfectly shelved
Zeeman state.  After state detection we depopulate the metastable $^2F_{7/2}$
state  using a pulse of 638\,nm light to re-enter the cooling cycle.

The resulting probabilities to find the population unshelved as a function of
shelving pulse length is shown in Fig.~\ref{fig:shelving} for the ion being
initially prepared in either $\ket{\downarrow}$ (red) or $\ket{\uparrow}$
(black).  We model the data using a rate equation that involves all relevant
levels and a saturation parameter of $s=0.12$, matching our observations. We
obtain a probability of 9(1)\,\% for the $\ket{\uparrow}$ state to be
off-resonantly shelved after $80\,\mu$s of shelving light. If the ion is
prepared in the $\ket{\downarrow}$ state, we find a probability of 28(3)\,\% to
remain unshelved.

To study the dynamics of the $^{171}$Yb$^+$ hyperfine states, we initialize the
ion in $F=0$ via optical pumping~\cite{Olmschenk:2007} and apply a microwave
pulse (rapid adiabatic passage) to prepare one of the three $F=1$ sublevels
before the interaction with the atoms. After the interaction, we measure the
population in the $F=1$ state by state-selective fluorescence imaging to obtain
a signal proportional to $\sum_{m_F}p_{\ket{1,m_F}}=1-p_{\ket{0,0}}$ as depicted
in Fig.~\ref{fig:levels} (right). To analyze the population in each of the
magnetic sublevels, we invert $p_{\ket{1,m_F}}$ with $p_{\ket{0,0}}$ by applying
a second microwave pulse before detection to get a signal proportional to
$1-p_{\ket{1,m_F}}$.

\section{Results}
\subsection{\yb{174}}
 \begin{figure}[t]
  \begin{center}
    \includegraphics[width=0.9\columnwidth]{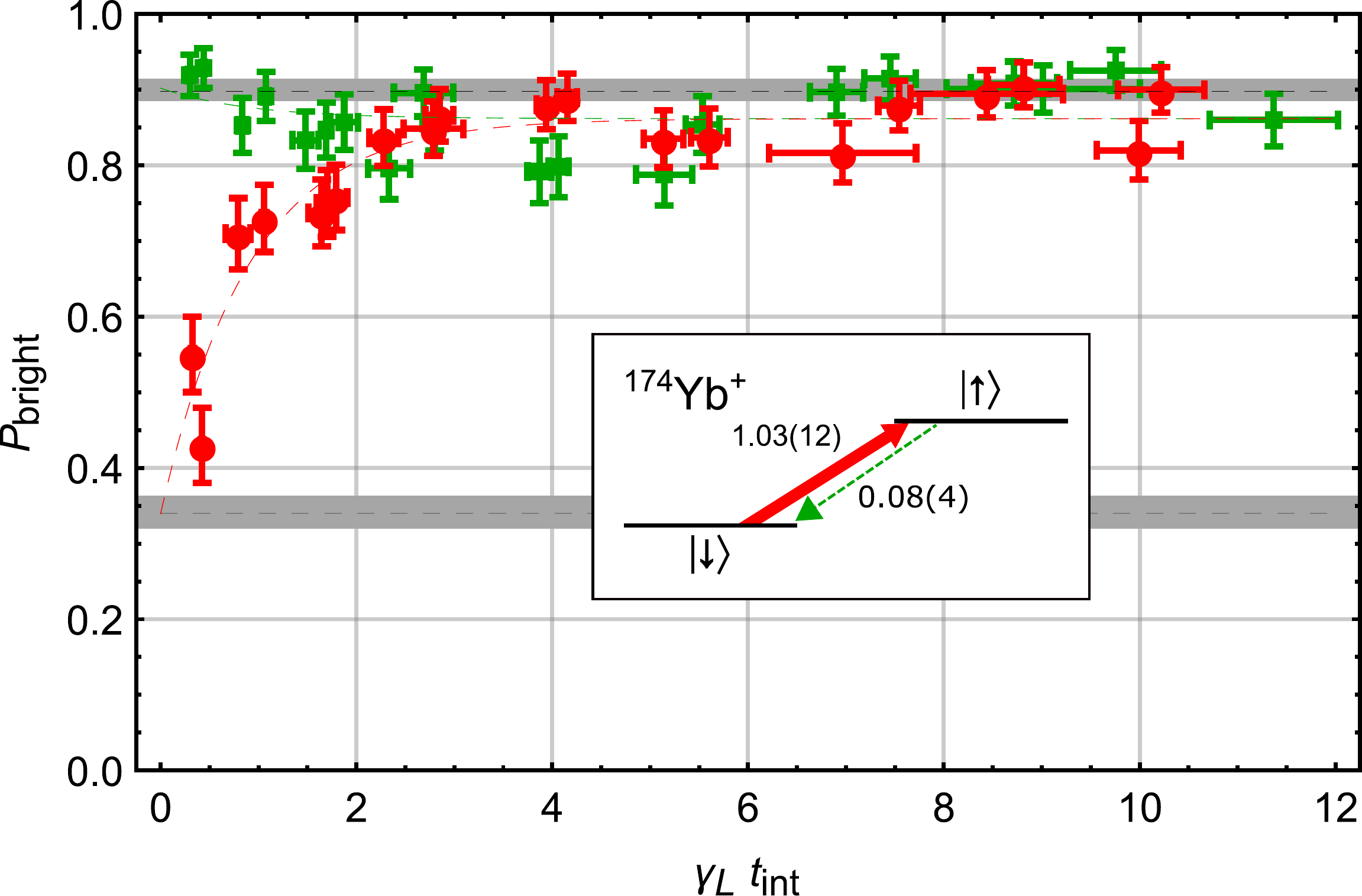}
  \end{center}
\caption{Probability for finding the $^{174}$Yb$^+$ ion in the bright state
   versus interaction time in units of the inverse Langevin rate. The gray bars
   indicate the minimum (for $\ket{\downarrow}$) and maximum (for
   $\ket{\uparrow}$) probabilities to find the ion in a bright state and their
   error range when no atoms are loaded, indicating the limits of the employed
   preparation and shelving techniques. The dashed lines are obtained from
   a combined two-level rate equation fit (Eq.~\ref{eqn:twolevel}). The inset
   shows the obtained spin flip rates of the two Zeeman sublevels in units of
   Langevin rates.}
  \label{fig:rates174}
\end{figure}
We scan the atom-ion interaction time $t_\text{int}$ in units of the inverse
Langevin rate $1/\gamma_\text{L}$ for $^{174}$Yb$^+$ initially prepared in one
of the two spin states.  The results are shown in Fig.~\ref{fig:rates174}. When
the ion is initialized in $\ket{\downarrow}$ (red discs), around one Langevin
collision is sufficient to flip its spin. In contrast, when initialized in the
$\ket{\uparrow}$ state (green squares), the ion keeps its initial polarization.
We fit the data to a two-level rate equation model~\cite{Ratschbacher:2013},
\begin{equation}
\begin{aligned}\label{eqn:twolevel}
P_{\text{b},\uparrow}(t_\text{int}) &=
  \left(P_{\text{b},\uparrow}^0-P_{\text{b}}^\infty\right)
  e^{-\gamma_\text{eq}t_\text{int}} + P_{\text{b}}^\infty\,,\\
P_{\text{b},\downarrow}(t_\text{int}) &=\left(P_{\text{b}}^\infty
  - P_{\text{b},\downarrow}^0\right)
  \left(1-e^{-\gamma_\text{eq}t_\text{int}}\right)
  + P_{\text{b},\downarrow}^0\,,
\end{aligned}
\end{equation}
where $P_{\text{b},m_J}^0$ are the probabilities to find an ion prepared in
$\ket{m_J}$ to be in the bright state when no atoms were loaded (lower and upper
gray bars in the plot), resembling the limits of our optical pumping and
shelving technique.  $P_{\text{b}}^\infty$ is the equilibrium probability to
appear bright after interaction with the atomic cloud. For the equilibration
rate we obtain $\gamma_\text{eq} = 1.1(1) \cdot\gamma_\text{L}$.  In the
two-level model $\gamma_\text{eq}=\gamma_+ + \gamma_-$, with $\gamma_\pm$ the
rates for $\Delta m_J = \pm 1$ transitions of the Zeeman state respectively.
From the control measurements (without atoms) we get $P_{\text{b},\downarrow}^0
= 0.34(2)$ and $P_{\text{b},\uparrow}^0 = 0.90(2)$.  Together with the
equilibrium probability $P_{\text{b}}^\infty = 0.86(1)$ we obtain the
equilibrium polarization of the ion $p_\uparrow^\infty= (P_{\text{b}}^\infty
-P_{\text{b},\downarrow}^0)/ (P_{\text{b},\uparrow}^0-P_{\text{b},\downarrow}^0)
= 0.93(4)$ as well as $\gamma_+ = 1.03(12)\cdot \gamma_\text{L}$ and $\gamma_-
= 0.08(4) \cdot\gamma_\text{L}$ from the relation $p_\uparrow^\infty
= \gamma_+/\left(\gamma_+ + \gamma_-\right)$.

\subsection{\yb{171}}
  \begin{figure*}[t]
  \begin{center}
    \includegraphics[width=1\textwidth]{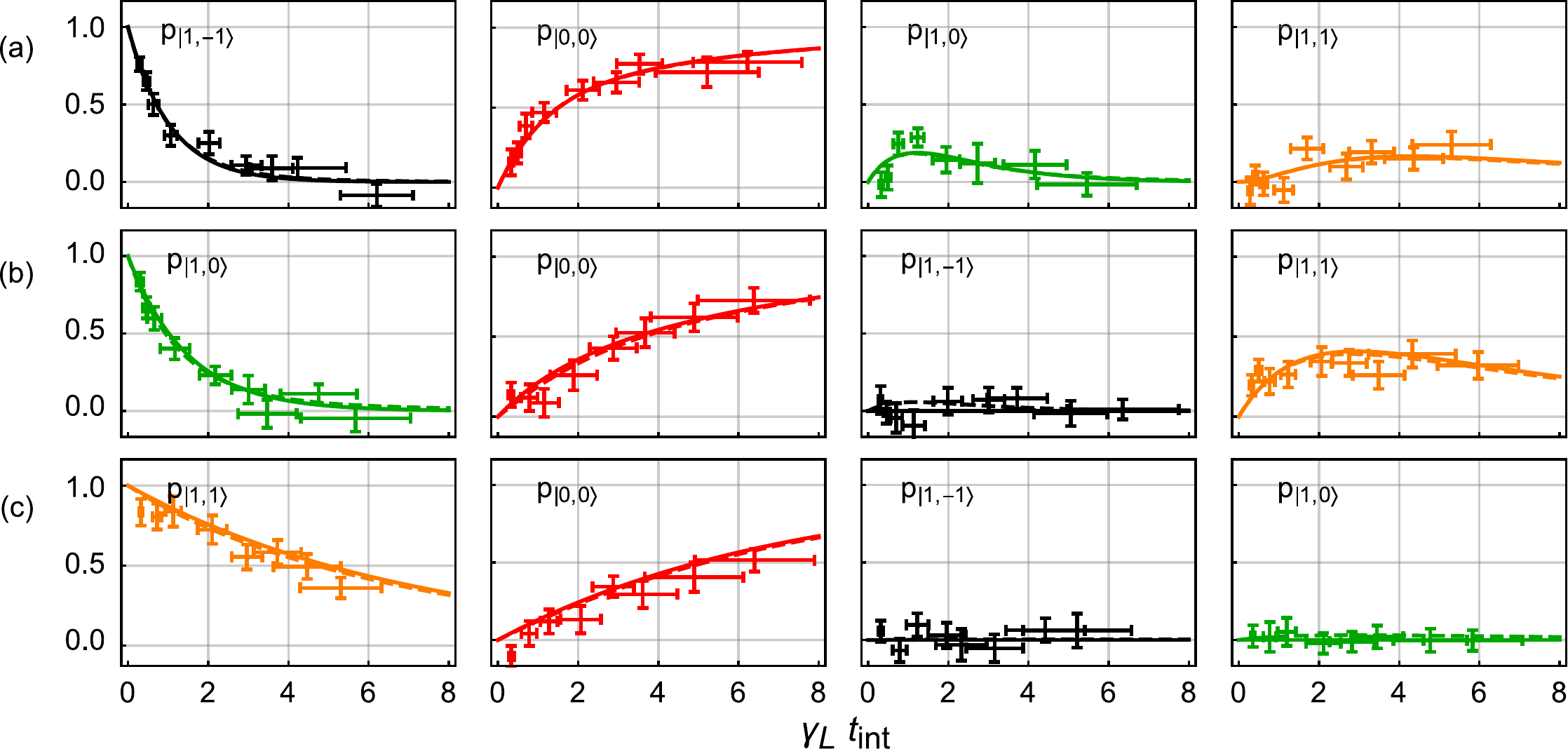}
  \end{center}
  \caption{Collision-induced population transfer in the $^{171}$Yb$^+$ hyperfine
  ground state after the preparation of $\ket{1,-1}$ (row (a)), $\ket{1,0}$ (row
  (b)) and $\ket{1,1}$ (row (c)). The first column shows the decay of
  the initially prepared state in the $F=1$ manifold and the second column the
  build up of population in the $F=0$ ground state. The other two columns show
  the population dynamics of two other states in $F=1$. The lines are obtained
  from a combined fit model assuming the rates for $\Delta m_F = -1$ within the
  $F=1$ manifold to be zero (solid) or allowing for all decay channels
  (dashed). The interaction time is given in units of the inverse Langevin rate.}
  \label{fig:allData}
\end{figure*}
The data obtained in the determination of population in the hyperfine states of
$^{171}$Yb$^+$ was fitted by the solutions of the four-level coupled rate
equations
\begin{align}
\begin{bmatrix}
{\dot{p}}_{\ket{0,0}}\\
{\dot{p}}_{\ket{1,-1}}\\
{\dot{p}}_{\ket{1,0}}\\
{\dot{p}}_{\ket{1,1}}
\end{bmatrix}
=
\hat{D}
\begin{bmatrix}
p_{\ket{0,0}}\\
p_{\ket{1,-1}}\\
p_{\ket{1,0}}\\
p_{\ket{1,1}}
\end{bmatrix}\,,
\label{eqn:fourlevels}
\end{align}
with the dot denoting the time derivative and the decay matrix
\begin{align}
\hat{D} =
\begin{bmatrix}
  0 & \Gamma_{\text{-}1} & \Gamma_{0} & \Gamma_{1} \\
  0 &-\Gamma_{\text{-}1}-\gamma_{\text{-}1,0} & \gamma_{0,\text{-}1} & 0 \\
  0 & \gamma_{\text{-}1,0} & -\Gamma_{0}-\gamma_{0,1} - \gamma_{0,\text{-}1} & \gamma_{1,0} \\
    0 & 0 & \gamma_{0,1} & -\Gamma_{1} - \gamma_{1,0}
\end{bmatrix}\,,
\nonumber
\end{align}
where $\Gamma_{m_F}$ denote the rates from $\ket{1,m_F}$ to $\ket{0,0}$ and
$\gamma_{m_F,m_F'}$ denote the rates from  $\ket{1,m_F}$ to $\ket{1,m_F'}$.
Note that we assumed the $\Delta m_F=\pm 2$ rates to be zero and that
transitions changing the total angular momentum by $\Delta F = +1$ are
energetically forbidden due to the 12.6~GHz hyperfine splitting. To obtain
analytic solutions of Eq.~\ref{eqn:fourlevels}, we need to set the
spin-nonconserving $\Delta m_F=-1$ rates $\gamma_{0,-1}$ and $\gamma_{1,0}$ to
be zero, since there was no clear evidence for these events in the experimental
data, as shown in the twelve relevant plots in Fig.~\ref{fig:allData}. To obtain
upper bounds on these two rates, we take the fitted curves (solid lines) as an
initial guess and numerically minimize the mean quadratic distance of the full
numerical solutions of Eq.~\ref{eqn:fourlevels} to the experimental data. The
optimized solutions are shown as dashed lines and deviate only slightly from the
initial guess.

\begin{figure}[t]
  \begin{center}
    \includegraphics[width=6.0cm]{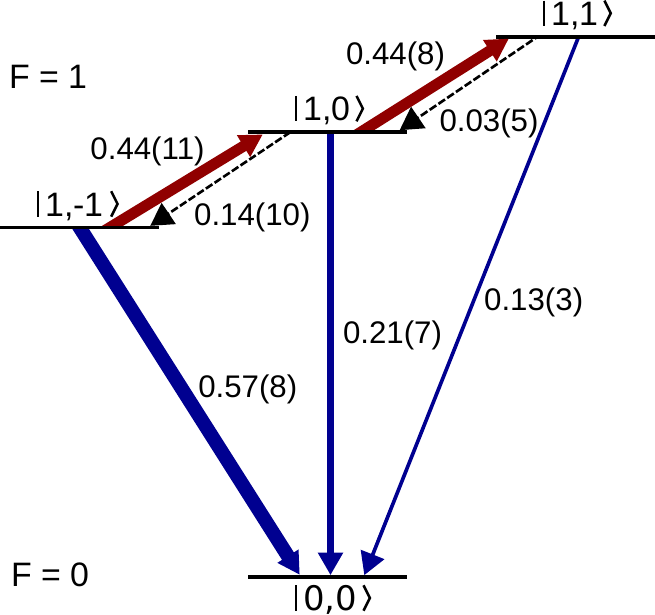}
  \end{center}
  \caption{Measured transition rates between the $^{171}$Yb$^+$ $^2$S$_{1/2}$
  hyperfine ground states in units of the Langevin rate $\gamma_\text{L}$. The
  $\Delta m_F=+1$ processes (dark red, pointing up right) dominate the dynamics,
  whereas the $\Delta m_F=-1$ transitions (dashed, black) can hardly be
  detected. The $\Delta F = -1$ transitions are shown
  in blue (pointing towards $F=0$).}
  \label{fig:rates171}
  \end{figure}

The resulting rates are shown in Fig.~\ref{fig:rates171}.  While the transition
rates $\gamma_{m_F,m_F'}$ for $\Delta{m_F} = m_F'-m_F=+1$ within the $F=1$
manifold are both approximately equal to $\gamma_{-1,0} = 0.44(11)\cdot
\gamma_\text{L} \approx \gamma_{0,1} = 0.44(8) \cdot\gamma_\text{L}$, the rates
$\Gamma_{m_F}$ changing the total angular momentum by $\Delta F = -1$ decrease
with increasing $m_F$ in the $F=1$ manifold from
$\Gamma_{-1}=0.57(8)\cdot\gamma_\text{L} $ via
$\Gamma_{0}=0.21(7)\cdot\gamma_\text{L}$ to
$\Gamma_{1}=0.13(3)\cdot\gamma_\text{L}$. Note that the decay $\Gamma_{1}$ does
not conserve the total spin of the atom-ion system. The rates changing only
$m_F$ by $-1$ are hardly detectable in our experiment due to the dominating
rates $\Gamma_{0}$ and $\Gamma_{-1}$.

\subsection{Purity of atomic spin}

\begin{figure}[b]
  \begin{center}
    \includegraphics[width=0.90\columnwidth]{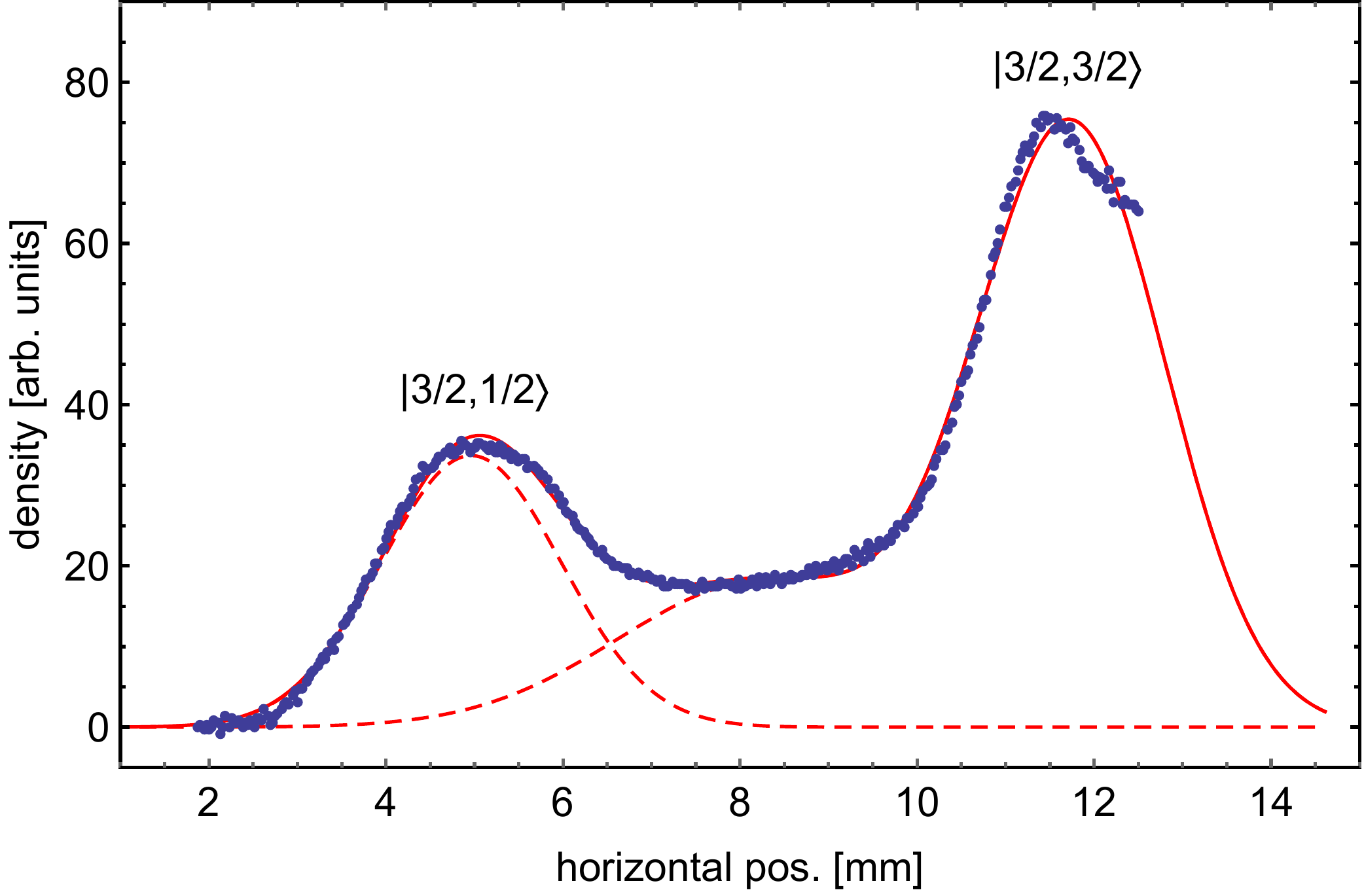}
  \end{center}
  \caption{Projected absorption image for a Stern-Gerlach acceleration time of
  $t_\text{SG} = 11.5$\,ms (blue points). The smaller peak on the left
  corresponds to the signal of the impurity atoms in the undesired
  $\ket{3/2,1/2}$ state lagging behind the atoms in $\ket{3/2,3/2}$ (right).
  Note that the x-axis points in the direction of acceleration. The data was fit
  using the sum of three Gaussian distributions (red, solid), one to model the
  impurity atoms (dashed, left) and two to model the majority atoms and the tail
  in between (dashed, right).}
  \label{fig:sg}
\end{figure}

The observed spin-nonconserving rates $\gamma_-$ and $\Gamma_{1}$ could be due
to second-order spin-orbit coupling that was recently suggested as a source of
spin-relaxation~\cite{Tscherbul:2016}. However, another possibility is that
atomic spin impurities within the gas cause sporadic collisions that appear as
spin non-conserving. In particular, we expect the presence of atoms in the
low-field seeking $\ket{3/2,1/2}$ state due to imperfect optical pumping. When
such impurity atoms collide with a spin-polarized ion, spin-allowed transitions
such as $\ket{3/2,1/2}_\text{atom}\ket{1,1}_\text{ion}\rightarrow
\ket{3/2,3/2}_\text{atom}\ket{0,0}_\text{ion}$ may occur that cannot be
distinguished from spin-relaxation caused by majority atoms. In this subsection,
we study the spin purity of the atomic cloud.

The spin of the atoms in the magnetic trap is polarized by applying
a $150\,\mu$s $\sigma^+$-polarized optical pumping pulse at the D1 transition
while applying an additional magnetic field of $1.0\,$mT along the beam
direction. Due to magnetic field inhomogeneities not every atom is in the
correct magnetic field to be pumped resonantly to the desired
$\ket{F=3/2,m_F=3/2}$
ground state. To estimate the purity of the magnetically trapped $^6$Li cloud in
the $\ket{3/2,3/2}$ state, we perform a Stern-Gerlach experiment. We
abruptly displace the trap minimum radially by approximately 6\,mm in less than
a millisecond while keeping the gradient constant at $g_{r} = 0.22\,$T/m. To
image the accelerating cloud, we switch off the trapping field after a variable
time $t_\text{SG}$. We wait 1\,ms for the magnetic fields to settle and take an
absorption image. After $t_\text{SG} = 11.5\,$ms the atoms in the
$\ket{3/2,1/2}$ state have separated from the main fraction in the
$\ket{3/2,3/2}$ state caused by the difference in magnetic moment, in agreement
with simulations. To make sure we image both fractions equally well, we scanned
the imaging laser frequency $\pm10$\,MHz around the resonance and average over
the results. In order to obtain the fraction of $\ket{3/2,1/2}$ atoms, we
project the images along the vertical axes, as it is shown in Fig.~\ref{fig:sg}.
The atoms in the $\ket{3/2,1/2}$ state (left peak) did not pass the trap minimum
at around 6\,mm yet whereas the atoms in the $\ket{3/2,3/2}$ state have reached
their turning point at around 12\,mm and show a long tail lagging behind, in
agreement with simulations. Due to the lack of a suitable model, we fit the data
with the sum of three Gaussian distributions (red, solid), one for the $\ket{3/2,1/2}$ state and two
for the $\ket{3/2,3/2}$ state to model the tail of the distribution (red,
dashed). By comparing the peak integrals, we obtain a fraction of
$\tilde{N}_{\ket{3/2,1/2}}=N_{\ket{3/2,1/2}}/N_\text{tot} = 24(1)\,\%$ of the
atoms being in the undesired state.

\begin{figure}[t]
  \begin{center}
    \includegraphics[width=0.90\columnwidth]{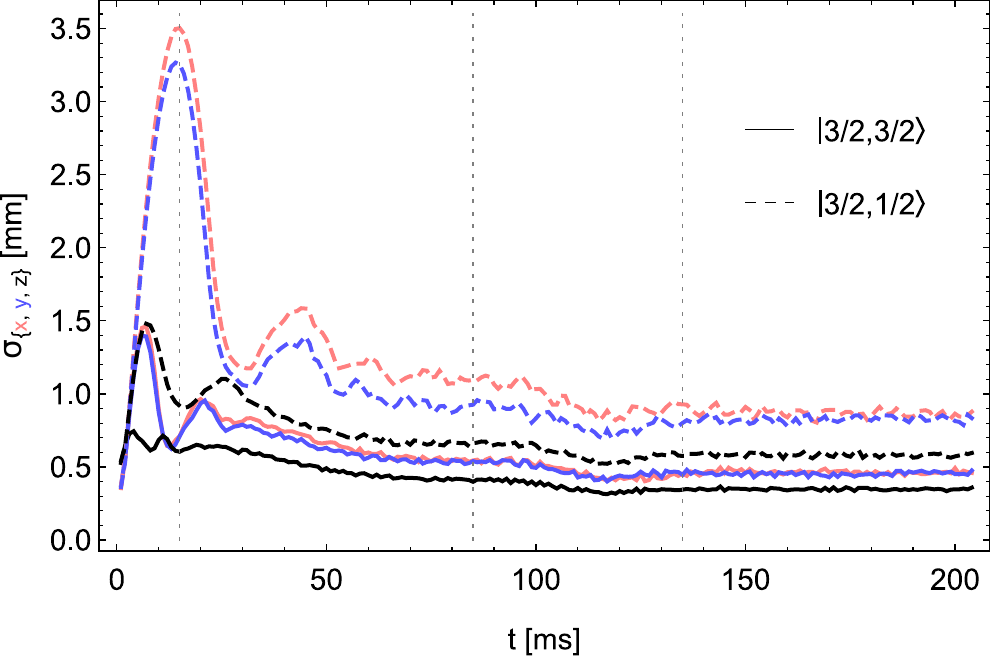}
  \end{center} \caption{Simulated evolution of the horizontal ($x$, light
  red/gray, $y$, blue/dark gray) and vertical ($z$, black) cloud sizes of
  $\ket{3/2,3/2}$ (solid) and $\ket{3/2,1/2}$ (dashed) for loading the magnetic
  trap from a compressed MOT (0-15\,ms) followed by a compressing transport to
  the interaction zone (the cloud is moved in the $x$-direction during the
  time-interval 15-85\,ms, and in the $z$-direction during the time-interval
  85-135\,ms). For more details, see
  Ref.~\cite{Joger:2017}.}
  \label{fig:sigmas}
\end{figure}

Due to the difference in magnetic moment, the spatial distribution for the
$\ket{3/2,1/2}$ state is expected to be broader than for the $\ket{3/2,3/2}$
state. Thus the possibility to find an impurity atom at the ion's position,
given by $\tilde{\rho}_{\ket{3/2,1/2}}= \rho_{\ket{3/2,1/2}}/
\rho_{\text{tot}}$, is reduced with respect to the fraction
$\tilde{N}_{\ket{3/2,1/2}}$. To estimate this ratio, we assume both fractions of
the cloud to have the same initial size and temperature before they are loaded
into the magnetic trap, justified by their origin from a compressed
magneto-optical trap (cMOT). When transferring the atoms from the cMOT to the
magnetic trap both temperature and size of the clouds change, depending on their
magnetic moment, proportional to their $m_F$ quantum number at low magnetic
fields. Using realistic parameters obtained from the experiment, we simulate
this transfer to the magnetic trap followed by a combined compression and
transport to the interaction zone within 135\,ms. The temporal evolution of the
cloud sizes is shown in Fig.~\ref{fig:sigmas}.
While the cloud size in $\ket{3/2,3/2}$ remains almost unchanged, the impurity
fraction initially expands because of its weaker trapping potential. Thus we
obtain a fractional density of $\tilde{\rho}_{\ket{3/2,1/2}} \leq 10\,\%$ at the
position of the ion.  Due to the increased cloud size, we assume to lose a large
fraction of the impurity atoms during the magnetic transport by collisions with
the ion trap electrodes such that the actual density fraction is expected to be
lower than in the simulation. Furthermore, spin-exchange with majority atoms
occurs, leading to the loss of the impurity atoms~\cite{Houbiers:1998}. We
justify these assumptions by performing the same Stern-Gerlach experiment as
described above, but on the atoms that return after the interaction time, where
we cannot observe an impurity spin signal anymore. However, since we cannot rule
out the presence of some impurity atoms at the location of the ion, we treat the
measured spin-nonconserving rates $\gamma_-$ and $\Gamma_{1}$ as an upper limit.

\section{Theory}
\label{Sec_Theory} To explain the measured rates, we construct and solve
a quantum microscopic model of cold atom-ion interactions and collisions based
on the \textit{ab initio} coupled-channel description of the Yb$^+$/Li system we
developed in Ref.~\cite{Tomza:2015}. As an entrance channel, we assume Li in
a spin-polarized state and Yb$^+$ in a selected state, while all allowed exit
channels are included in the model.
\begin{figure}[tb!]
	\centering
	\includegraphics[width=0.90\columnwidth]{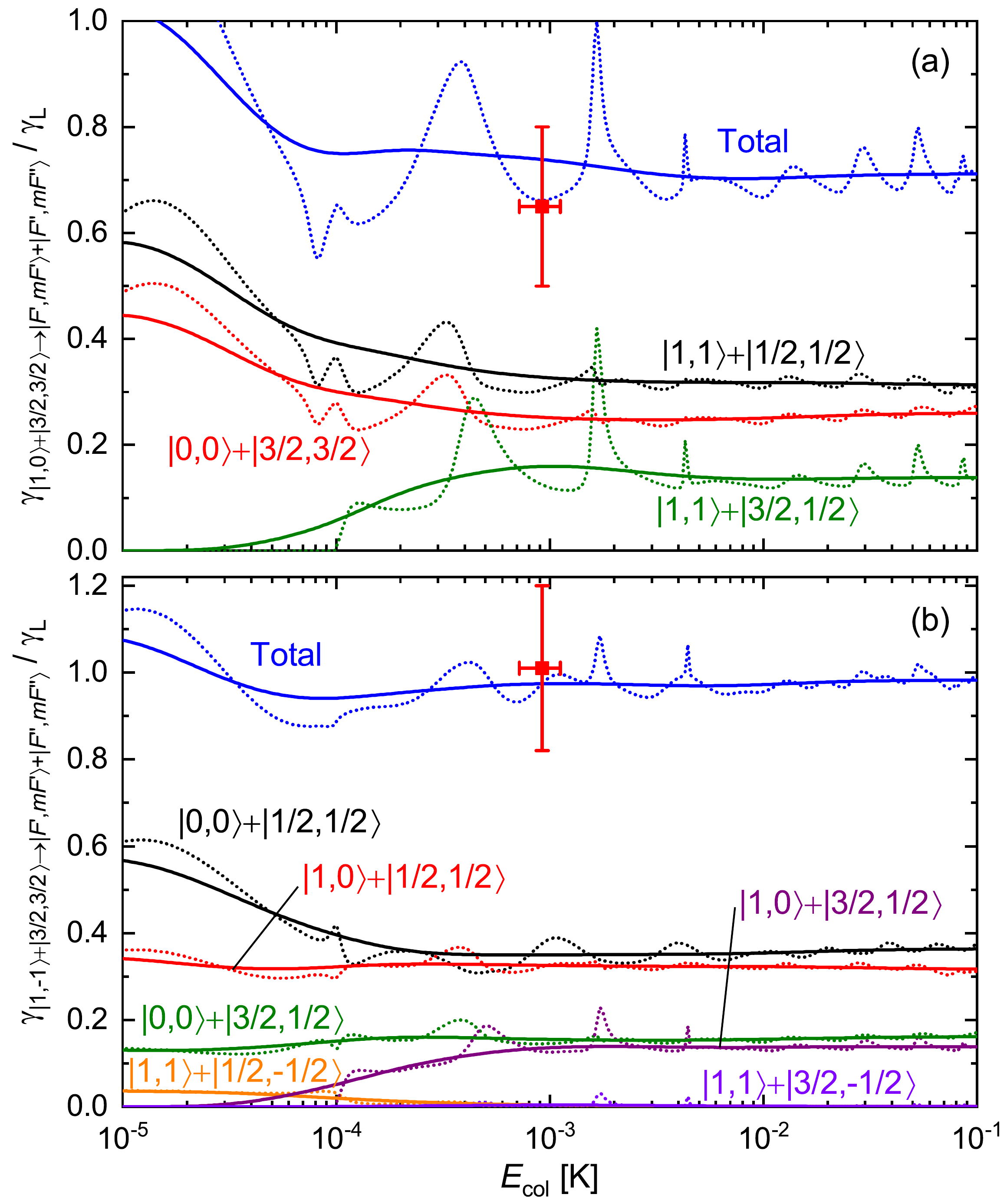}
  \caption{Total spin-exchange rates and their decompositions onto all possible
  spin states for $^{171}$Yb$^+$/$^6$Li versus collision energy obtained in
  coupled-channel scattering calculations compared with measured rates. Panel
  (a) for the $|1,0\rangle$ state and (b) for the $|1,-1\rangle$ state of the
  $^{171}$Yb$^+$ ion. Triplet and singlet scattering lengths of
  $a_\text{T}=-a_\text{S}=R_4$ are assumed. Dotted lines are energy-resolved
  rates and solid lines are thermally averaged rates.}
	\label{fig:171Yb}
\end{figure}

\subsection{Spin exchange rates}
To explain the observed spin-exchange rates, we perform \textit{ab initio}
quantum scattering calculations as implemented in
Refs.~\cite{Tomza:2015,Tomza:2014}. The Hamiltonian describing the nuclear
motion of the Yb$^+$/Li atom-ion system reads
\begin{equation}\label{eq:Ham}
\begin{split}
    \hat{H}=&-\frac{\hbar^2}{2\mu}\frac{1}{R}\frac{d^2}{dR^2}R+
    \frac{\hat{l}^2}{2\mu R^2}+
    \sum_{S,M_S}V_S(R)|S,M_S\rangle\langle S,M_S|\\
    &+\hat{H}_{\mathrm{dip}}+\hat{H}_{\mathrm{Yb}^+}+\hat{H}_{\mathrm{Li}}\,,
\end{split}
\end{equation}
where $R$ is the atom-ion distance, $\hat{l}$ is the rotational angular momentum
operator, $\mu$ is the reduced mass, and $V_S(R)$ is the potential energy curve
for the state with total electronic spin $S$. $\hat{H}_{\mathrm{dip}}$ is the
effective dipolar-like interaction. The atomic Hamiltonian, $\hat{H}_j$
($j=$Yb$^+$, Li), including hyperfine and Zeeman interactions, is given by
\begin{equation}\label{eq:Ham_at}
\hat{H}_j=\zeta_{j}\hat{i}_{j}\cdot\hat{s}_{j}
  +\left(g_e\mu_{{\rm B}}\hat{s}_{j,z}+g_{j}\mu_{{\rm N}}\hat{i}_{j,z}\right)B_z\,,
\end{equation}
where $\hat{s}_{j}$ and $\hat{i}_{j}$ are the electron and nuclear spin
operators, $\zeta_{j}$ is the hyperfine coupling constant, $g_{e/j}$ is the
electron/nuclear $g$ factor, and $\mu_{\text{B/N}}$ is the Bohr/nuclear
magneton, respectively.  For the fermionic $^{174}$Yb$^+$ ion, Eq.~\eqref{eq:Ham_at} reduces
to the electronic Zeeman term. A magnetic field of 0.42$\,$mT is assumed in all
calculations, as used in the experiment.

We use potential energy curves for the $a^3\Sigma^+$ and $A^1\Sigma^+$
electronic states as calculated in Ref.~\cite{Tomza:2015}. The scattering
lengths are fixed by applying uniform scaling factors $\lambda_S$ to the
interaction potentials: $V_S(R)\to\lambda_S V_S(R)$. We express scattering
lengths in units of the characteristic length scale for the ion-atom interaction
$R_4=\sqrt{2\mu C_4/\hbar}$.

We construct the total scattering wave function in a fully uncoupled basis set,
\[
|i_\mathrm{Yb^+},m_{i,\mathrm{Yb^+}}\rangle
|s_\mathrm{Yb^+},m_{s,\mathrm{Yb^+}}\rangle
|i_\mathrm{Li},m_{i,\mathrm{Li}}\rangle
|s_\mathrm{Li},m_{s,\mathrm{Li}}\rangle|l,m_l\rangle\,,
\]
where $m_j$ is the projection of angular momentum $j$ on the space-fixed $z$-axis,
assuming the projection of the total angular momentum
$M_\mathrm{tot}=m_{f,\mathrm{Yb^+}}+m_{f,\mathrm{Li}} + m_l=m_{i,\mathrm{Yb^+}}
+ m_{s,\mathrm{Yb^+}}+m_{i,\mathrm{Li}}+m_{s,\mathrm{Li}}+m_l$
to be conserved.
We solve the coupled-channels equations using a renormalized Numerov
propagator~\cite{Johnson:1978} with step-size doubling and about 100 step points
per de Broglie wavelength. The wave function ratio $\Psi_{i+1}/\Psi_{i}$ at the
$i$-th grid step is propagated to large interatomic separations, transformed to
the diagonal basis, and the $K$ and $S$ matrices are extracted by imposing
long-range scattering boundary conditions in terms of Bessel functions.
\begin{figure}[tb]
\begin{center}
\includegraphics[width=1\columnwidth]{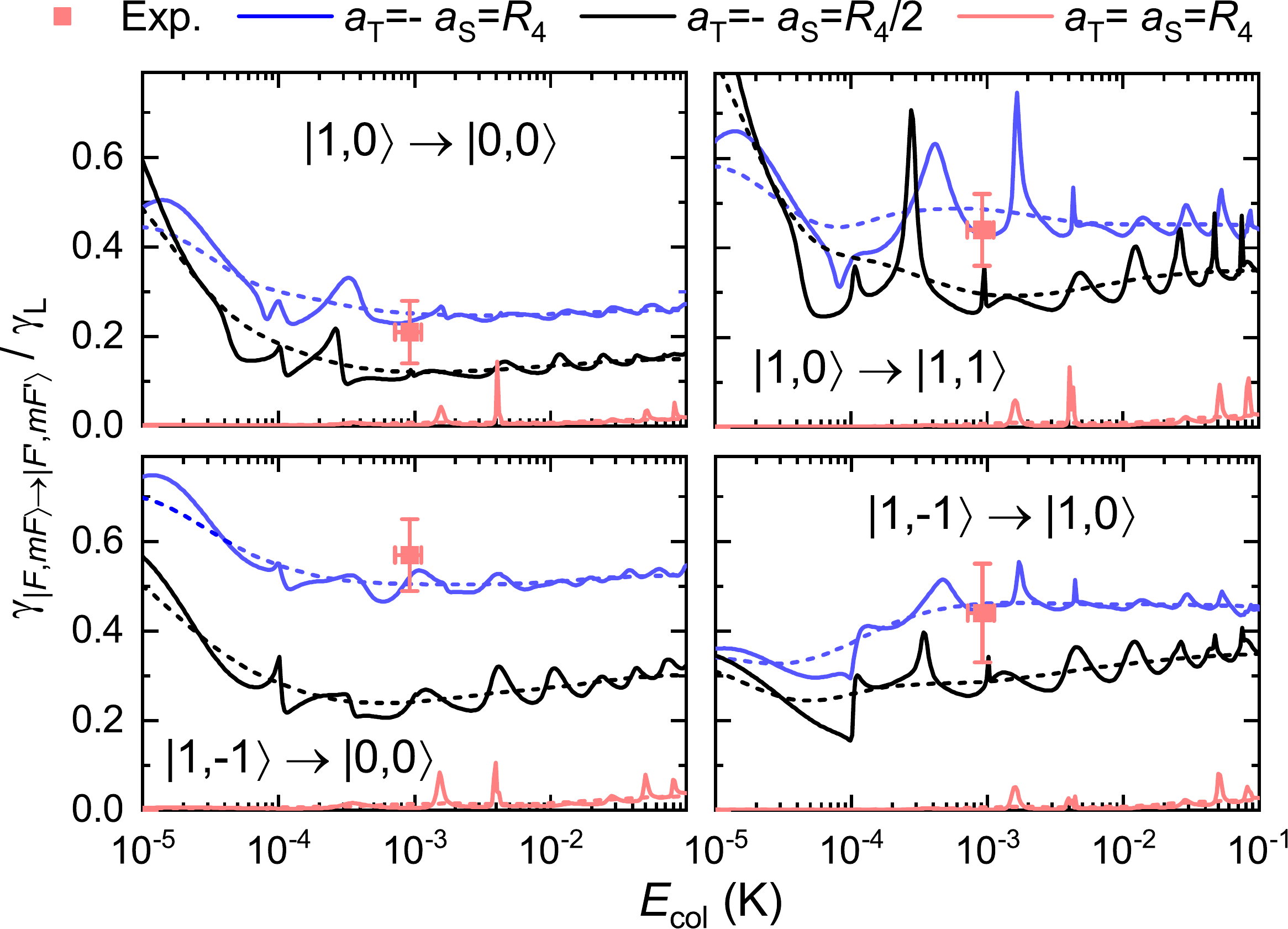}
\end{center}
\caption{Spin-exchange transition rates for $^{171}$Yb$^+$ versus collision
  energy obtained in coupled-channel scattering calculations for three sets of
  scattering lengths compared with measured rates (red/gray squares). Dotted
  lines are energy-resolved rates and solid lines are thermally averaged rates.
  The lowest rates are obtained for $a_\text{T}=a_\text{S}=R_4$ (light red/gray) and the
  highest for $a_\text{T}=-a_\text{S}=R_4$ (blue/dark gray) within the three
  sets shown over a broad temperature range.}
\label{fig:theory}
\end{figure}

We calculate elastic $K^{i}_{\rm el}(E)$ and inelastic $K^{i}_{\rm in}(E)$ rate
constants for collisions in the $i$-th channel from the diagonal elements of the
$S$ matrix summed over partial waves~$l$,
\begin{equation}
\begin{split}
K^{i}_{\rm el}(E) &= \frac{\pi\hbar}{\mu
  k_i}\sum_{l=0}^\infty(2l+1)\left|1-S_{ii}^l(E)\right|^2\,,\\
K^{i}_{\rm in}(E) &= \frac{\pi\hbar}{\mu
  k_i}\sum_{l=0}^\infty(2l+1)\left(1-|S_{ii}^l(E)|^2\right)\,,
\end{split}
\end{equation}
and state-to-state inelastic $K^{ij}_{\rm in}(E)$ rate constants for the
transition between $i$-th and $j$-th channels from the off-diagonal elements of
the $S$ matrix summed over partial waves~$l$,
\begin{equation}\label{eq:Kij}
K^{ij}_{\rm in}(E) = \frac{\pi\hbar}{\mu
  k_i}\sum_{l=0}^\infty(2l+1)\left|S_{ij}^l(E)\right|^2\,.
\end{equation}
Here $k_i=\sqrt{2\mu (E-E_i^\infty)/\hbar^2}$ is the $i$-th channel wave vector
with $E$ the collision energy and $E_i^\infty$ the $i$-th threshold energy.

The spin-exchange rates between different states of the Yb$^+$ ion colliding
with Li atoms are calculated as a sum over all possible state-to-state
transitions obtained with Eq.~\eqref{eq:Kij}. The exemplary decompositions of
the total spin-exchange rates for the $|1,0\rangle$ and $|1,-1\rangle$ states of
the $^{171}$Yb$^+$ ion colliding with $|3/2,3/2\rangle$ state $^6$Li atoms are
presented in Fig.~\ref{fig:171Yb}. The opening of relevant spin channels at an
energy of 0.1$\,$mK can be seen. These calculations also confirm that the $\Delta
m_F=+2$ transitions are negligible at our experimental conditions.

Spin-exchange rates depend on the difference between
singlet and triplet scattering phases~\cite{Julienne:2010}.
Results for the $^{171}$Yb$^+$ ion are presented in Fig.~\ref{fig:theory}. It
can be seen that even in the mK temperature regime the spin-exchange rates still
depend strongly on the difference between assumed singlet ($a_\text{S}$) and
triplet ($a_\text{T}$) scattering lengths~\cite{Sikorsky:2018lock,Cote:2018}.
To reproduce
the large spin-exchange rates measured, the difference between the scattering
lengths has to be close to the one that maximizes the scattering phase
difference, that is $|a_\text{T}-a_\text{S}|=2 R_4$, where
$R_4=1319\,$bohr for
Yb$^+$/$^6$Li. Similar results are found for the $^{174}$Yb$^+$ ion. If we take
into account that part
of the measured rates may be due to spin-nonconserving transitions, the
difference between singlet and triplet scattering lengths is still restricted to
a large value $R_4<|a_\text{T}-a_\text{S}|<3R_4$, assuming the largest measured
values of the spin-nonconserving rates. The large difference between singlet and
triplet scattering length also has another meaningful consequence, namely that
broad magnetic Feshbach resonances can be expected when the $s$-wave regime of
collisions is reached~\cite{Tomza:2015,Julienne:2010}.

\subsection{Magnetically tunable Feshbach resonances}

To assess the prospects for the observation and application of magnetically
tunable Feshbach resonances in cold Yb$^+$/$^6$Li collisions, we calculate
thermally averaged rates for elastic and inelastic collisions as a function of
a magnetic field between 0 and 1000 Gauss. We consider several possible entrance
channels of the $^{174}$Yb$^+$/$^6$Li and $^{171}$Yb$^+$/$^6$Li systems at
collision energies of $E_{\text{col}}/k_\text{B}=100\,$nK, 10$\,\mu$K and
1$\,$mK. We assume the difference between the singlet and triplet scattering
lengths to be close to the one that maximizes the scattering phase difference,
that is $|a_\text{T}-a_\text{S}|=2 R_4$, to reproduce the observed large rates
for spin-exchange collisions.

\begin{figure}[tb!]
	\centering
	\includegraphics[width=0.90\columnwidth]{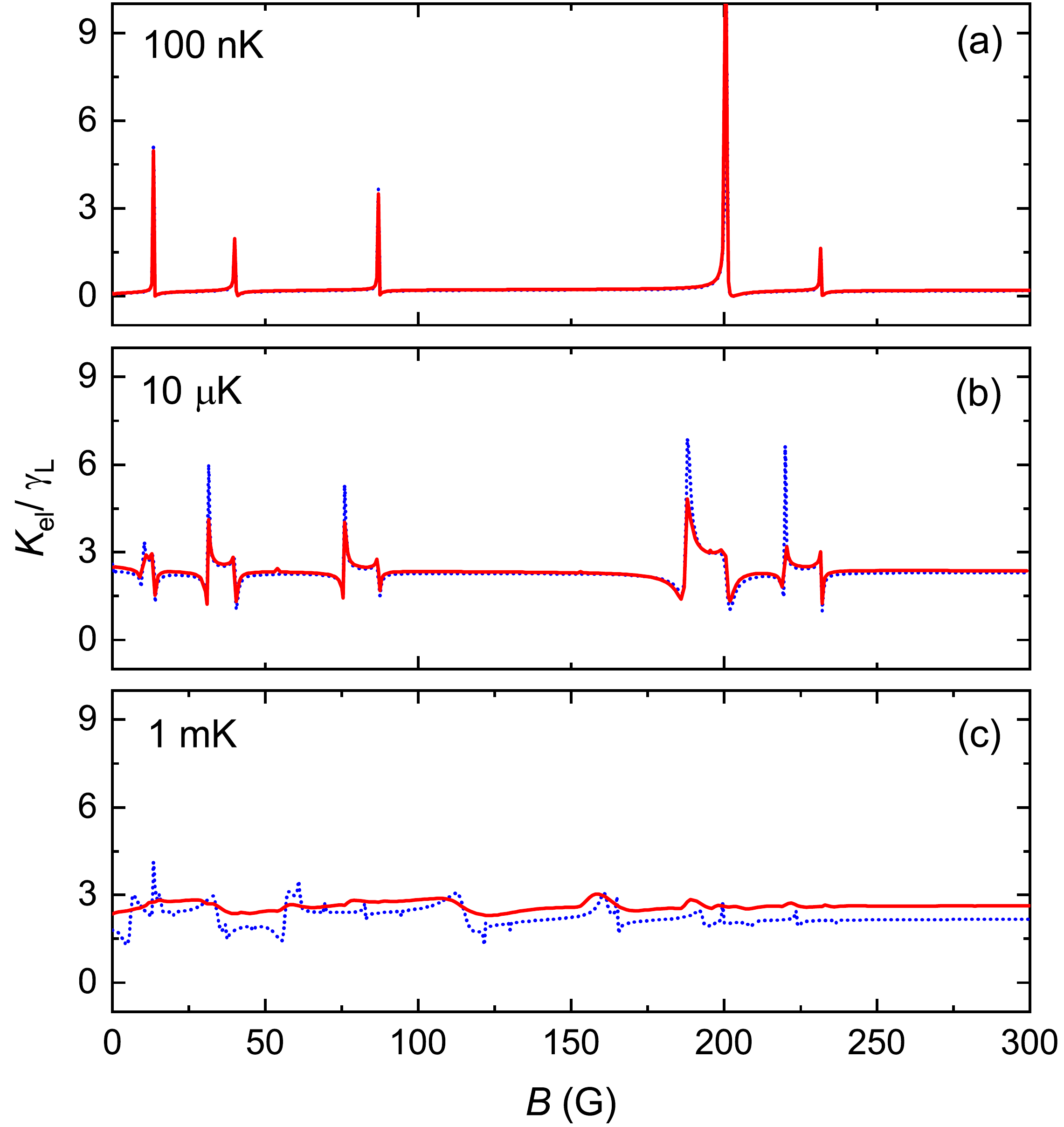}
  \caption{Elastic scattering rates for $^{174}$Yb$^+$/$^6$Li versus magnetic
  field for the lowest channel with
  $M_F=m_{f,\mathrm{Yb^+}}+m_{f,\mathrm{Li}}=0$ at
  (a)~$E_{\text{col}}/k_\text{B}= 100\,$nK, (b) 10$\,\mu$K, and (c) 1$\,$mK. The
  singlet scattering length is set to $a_\text{S}=-R_4$ and the triplet
  scattering length to $a_\text{T}=R_4$. Blue dotted lines are energy-resolved
  rates and red solid lines are thermally averaged rates.}
	\label{fig:FR}
\end{figure}

An example result for $^{174}$Yb$^+$/$^6$Li is presented in Fig.~\ref{fig:FR}.
Based on our detailed analysis, we can draw
some general conclusions. First of all, for the assumed difference of the
singlet and triplet scattering lengths, several measurable resonances are
expected for experimentally accessible magnetic fields below 500$\,$G for most
values of $M_\text{tot}$ at the energetically lowest channels. For higher energy
channels, the resonances start to be less pronounced due to possible
spin-exchange losses. At the same time, however, magnetic resonances can be used
to control spin-exchange rates to a large extent. For spin-polarized collisions
or very energetic channels no Feshbach resonances are expected. The
temperature-dependence visible in Fig.~\ref{fig:FR} is typical for all
combinations investigated. For collision energies deep in the quantum regime
$E_{\text{col}}=k_\text{B}\cdot 100\,$nK the Feshbach resonances are very
pronounced. For $E_{\text{col}}=k_\text{B}\cdot 1\,$mK they are not present
because of the contribution of many partial waves and thermal averaging. Finally, for
$E_{\text{col}}=k_\text{B}\cdot 10\,\mu$K, which should be in reach in the
present system~\cite{Fuerst:2018}, elastic collisions are dominated by
contributions from $s$- and $p$-waves only, and broad and measurable resonances
can be expected.

\subsection{Spin-relaxation}

The spin-nonconserving relaxation is governed by the effective Hamiltonian
describing dipolar-like interaction between the electronic spins of the Yb$^+$
ion, $\hat{{s}}_\text{Yb$^+$}$, and the Li atom,
$\hat{{s}}_\text{Li}$~\cite{Mies:1996,Tscherbul:2016}
\begin{equation}\label{H}
\hat{H}_\text{dip} = \left(-\frac{\alpha^2}{R^3}
  + \lambda_\text{SO}(R) \right) \left[3\hat{s}^z_\text{Yb$^+$}
  \hat{s}^z_{\text{Li}} - \hat{{s}}_\text{Yb$^+$} \cdot
  \hat{{s}}_\text{Li}\right]\,,
\end{equation}
where $\alpha$ is the fine-structure constant. The first term
$-\frac{\alpha^2}{R^3}$ describes the contribution due to the direct magnetic
dipole-dipole interaction. The second and dominating term $\lambda_\text{SO}(R)$
describes the effective dipole-dipole interaction in the second order of
perturbation theory due to the first-order spin-orbit couplings between the
$a^3\Sigma^+$ triplet electronic ground state and $^3\Pi$ electronic excited
states. This interaction was identified as the main source of spin-nonconserving
relaxation in the Yb$^+$/Rb system~\cite{Ratschbacher:2013,Tscherbul:2016}.

The spin-orbit coupling coefficient $\lambda_\text{SO}(R)$ can be calculated
from the energy difference between the $a0^-$ and $a1$ relativistic electronic
states~\cite{Tscherbul:2016} or using second-order perturbation theory with
non-relativistic electronic states and matrix elements of the spin-orbit
coupling Hamiltonian between them~\cite{Mies:1996}. Using the latter approach, we
calculate the spin-orbit coupling coefficient given by
\begin{equation}\label{eq:SO}
\lambda_\text{SO}(R)=\frac{2}{3}\frac{|\langle a^3\Sigma^+ |H_\text{SO}|b^3\Pi
  \rangle|^2}{V_{b^3\Pi }(R)-V_{a^3\Sigma^+}(R)}\,,
\end{equation}
where $\langle a^3\Sigma^+ |H_\text{SO}|b^3\Pi \rangle$ is the matrix element of
the spin-orbit coupling between the $a^3\Sigma^+$ and $b^3\Pi$ electronic
states. $V_{b^3\Pi }(R)$ and $V_{a^3\Sigma^+}(R)$ are potential energy curves of
the $a^3\Sigma^+$ and $b^3\Pi$ electronic states which we calculated accurately
in Ref.~\cite{Tomza:2015}. In Eq.~\eqref{eq:SO} we neglected terms originating
from the coupling between the $a^3\Sigma^+$ triplet ground state and higher
$^3\Pi$ states because these terms should be much smaller due to a smaller
spin-orbit coupling in the numerator and a larger energy difference in the
denominator. The matrix elements of the spin-orbit coupling Hamiltonian,
$H_\text{SO}$, are evaluated using wave functions calculated using \textit{ab
initio} electronic structure methods from Ref.~\cite{Tomza:2012}.

\begin{figure}[t]
	\centering
	\includegraphics[width=0.90\columnwidth]{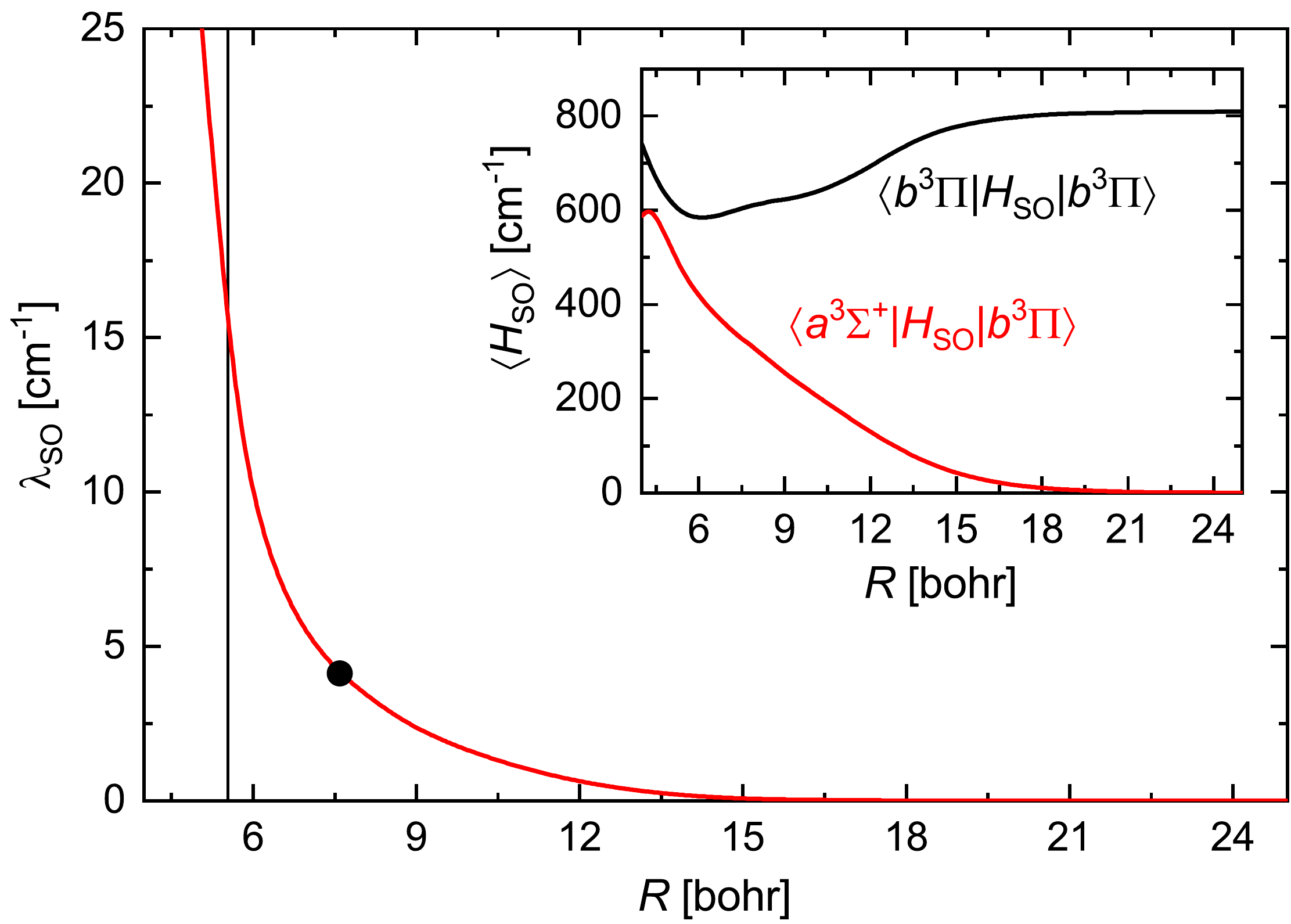}
  \caption{Second-order spin-orbit coupling coefficient $\lambda_\text{SO}(R)$
  for the Yb$^+$/Li system (red/gray line) as a function of the atom-ion
  distance. The point and vertical line indicate the value for the equilibrium
  distance and the position of the classical turning point of the $a^3\Sigma^+$
  electronic state, respectively. The inset shows the matrix elements of the
  spin-orbit coupling for the $a^3\Sigma^+$ (red/gray) and $b^3\Pi$ (black)
  electronic states of the Yb$^+$/Li system.}
	\label{fig:SO}
\end{figure}

The matrix elements of the spin-orbit coupling for the $a^3\Sigma^+$ and
$b^3\Pi$ electronic states are presented in the inset of Fig.~\ref{fig:SO}. The
matrix element of the spin-orbit coupling for the $b^3\Pi$ electronic state
$\langle b^3\Pi  |H_\text{SO}|b^3\Pi \rangle$ asymptotically reaches the value
which reproduces the experimental spin-orbit splitting of the $^3P$ state of the
Yb atom~\cite{nist}. This confirms the quality of our calculations. The matrix
element of the spin-orbit coupling between the $a^3\Sigma^+$ and $b^3\Pi$
electronic states $\langle a^3\Sigma^+ |H_\text{SO}|b^3\Pi \rangle$ decreases
exponentially with the atom-ion distance as expected but at the equilibrium
distance of the $a^3\Sigma^+$ electronic state it still has a significant value
of 323$\,$cm$^{-1}$.

The calculated second-order spin-orbit coupling coefficient
$\lambda_\text{SO}(R)$ is presented in Fig.~\ref{fig:SO}. It decreases
exponentially with the atom-ion distance but at the equilibrium distance and
classical turning point of the $a^3\Sigma^+$ electronic state it has a value of
4.1$\,$cm$^{-1}$ and 14.2$\,$cm$^{-1}$, respectively. These values are an order
of magnitude smaller than for the Yb$^+$/Rb
system~\cite{Tscherbul:2016,Sayfutyarova:2013} and an order of magnitude larger
than for neutral alkali-metal atoms~\cite{Mies:1996}.  The larger second-order
spin-orbit coupling coefficient for the Yb$^+$/Rb system is due to the small
energy difference and crossing between the $a^3\Sigma^+$ and $b^3\Pi$ electronic
states and a contribution from the Rb atom to the spin-orbit interaction which
is smaller for the very light Li atom. The spin-orbit coupling for neutral
alkali-metal atoms is smaller because relativistic effects are smaller for
alkali-metal atoms as compared to the heavy Yb$^+$ ion.

\section{Conclusions}
We have measured the spin dynamics of single trapped Yb$^+$ ions immersed in
a cold cloud of spin-polarized $^6$Li atoms. This combination is of significant
interest as its large mass ratio may allow it to reach the quantum regime in
Paul traps~\cite{Cetina:2012,Fuerst:2018}. We have observed very fast
spin-exchange that occurs within a few Langevin collisions.  Spin-relaxation
rates are found to be a factor~$\geq$~13(7) smaller than spin-exchange rates in
$^{174}$Yb$^+$. Spin impurity atoms in the atomic cloud may lead to apparent
spin-relaxation, such that we interpret the observed relaxation rate as an upper
limit.  The observed ratio between spin-allowed and spin-nonconserving
collisions is higher than those observed in Yb$^+$/Rb~\cite{Ratschbacher:2013},
where a ratio of $0.56(8)$ was measured for Rb atoms in the stretched
$\ket{F=2,m_F=2}$ state. For Sr$^+$/Rb~\cite{Sikorsky:2018}, both spin-exchange
and spin-relaxation rates for Rb atoms prepared in the $\ket{F=1,m_F=-1}$ state
are lower than the rates observed in this work and have a lower ratio of
$5.2(8)$.  For $^{171}$Yb$^+$, we have measured the decay channels of all spin
states within the ground state hyperfine manifold and observe both spin-exchange
and spin-relaxation processes. We have compared our measured rates to
predictions from \textit{ab initio} electronic structure and quantum scattering
calculations and conclude that a large difference between singlet and triplet
scattering lengths is responsible for the observed large spin-exchange rates,
whereas small second-order spin-orbit coupling results in small spin-relaxation
rates. These findings suggest good prospects for the observation of Feshbach
resonances in the Yb$^+$/Li system.

\section*{Acknowledgements}
This work was supported by the European Union via the European Research Council
(Starting Grant 337638) and the Netherlands Organization for Scientific Research
(Vidi Grant 680-47-538 and Start-up grant 740.018.008) (R.G.). M.T.~was
supported by the National Science Centre Poland (Opus Grant~2016/23/B/ST4/03231)
and PL-Grid Infrastructure.

\section*{References}
%
\end{document}